# On the Question of the Behavior of O–C Residuals of the Active Algol-Like Binary RZ Cassiopeiae

**Alex V. Golovin**
*Kyiv National Shevchenko University, Physics Department; Visiting astronomer, Crimean Astrophysical Observatory, Berdyansk, Crimea, Ukraine*
*astronom_2003@mail.ru*

**Elena P. Pavlenko**
*Crimean Astrophysical Observatory, Nauchny, Crimea, Ukraine*
*pavlenko@crao.crimea.ua*

*Received February 4, 2005; revised May 16, 2005; accepted June 8, 2005*

**Abstract** We present an analysis of the RZ Cassiopeia O–C diagram for a 30-year period, based on data published by the AAVSO. The new parabolic light-elements are determined and $\dot{P}$ was estimated. This enabled us to establish the value of $\dot{M}$, the rate of the mass transfer between components of this eclipsing binary. Deviations from the parabola in the RZ Cas O–C diagram have a wavelike trend. The pulsation of the primary component influences the scatter of O–C points, as well as it influences the shape of the eclipse. Some physical points on such O–C behavior are presented.

## 1. Introduction

Orbital period changes in Algol-like eclipsing binary stars are interesting phenomena. This paper examines the period of the active Algol-like system RZ Cas (= TYC4317-1793-1 = HD 17138 = SAO 12445 = HIP 13133 = GSC 04317-01793 = BD +69 179 = 1 RXS J024854.7+693804 = AN 77.1906) that has exhibited large period changes over the last 30 years of intensive visual observations. Our goals are: 1) to illustrate the phenomenon by looking at some recent period changes revealed by the observations of a great number of authors; 2) to provide an accurate and up-to-date ephemeris for this active eclipsing binary star; and 3) to attempt to explain any changes in the O–C (observed-minus-calculated) diagram.

An O–C diagram for RZ Cas was generated and analyzed. This diagram shows the difference between the observed time of minimum and the time calculated by a formula that assumed a constant period.

The A3V+K0 IV eclipsing binary RZ Cas is an active semi-detached Algol system showing complex features in its light-curve. The variability of RZ Cas was discovered by G. Müller (1906). The primary minimum is a partial eclipse (Narusawa *et al.* 1994) and the light-curve is distorted, possibly, by star spots (e.g. Maxted *et al.* 1994) and/or by non-symmetric circumstellar matter (Olson 1982; Varricatt *et al.* 1998). The non-symmetric distribution of circumstellar matter was also suggested by Richards and Albright (1999) based on the observation of single-peak emission.

Olson (1982) selected five Algol-type semidetached close binaries which show strong activity, and called them *active Algols*. They are: U Cep, RW Tau, U CrB, U Sge, and our RZ Cas. He postulated the existence of a small accretion disk around the more massive primary component and a hot spot on the disk. As in the case of another active system, U Cep, he argued that an accretion disk with hot and cool regions could explain the observed phenomena of RZ Cas. Duerbeck and Hanel (1979) made spectroscopic observations of RZ Cas and obtained the spectral classification A3V for the primary component and the mass function $f(M) = 0.043 \pm 0.003\, M_{\odot}$ for the system. In that work the peculiar behavior of the radial-velocity curve of the system was found, which was probably due to the existence of circumstellar matter.

Strong activity in this Algol has been detected at radio and X-ray wavelengths. For example, McCluskey and Kondo (1984) observed several close binaries including RZ Cas with the help of the Einstein Observatory. They found RZ Cas to be a fairly strong X-ray source with the X-ray luminosity $3.92 \times 10^{30}$ erg s$^{-1}$ at the phase of $0\overset{p}{.}54$. This may prove that the chromosphere of the secondary is strongly active in the RZ Cas system. Drake *et al.* (1986) have made radio observations of 13 binaries with VLA at 6 cm and detected emission from eight systems, of which seven are considered to be RS CVn-like systems and the eighth is RZ Cas. The flux density (1.4 mJy) of RZ Cas is comparable to those of the short period RS CVn binaries and its radio spectrum is also similar to those of RS CVn-like systems. Umana *et al.* (1991) have measured 14 Algol-like systems with the VLA at 6 cm and detected radio emission from seven systems including RZ Cas. The radio flux density of RZ Cas changed from 3.25 mJy (1984 November 9) to 1.25 mJy (1989 February 18). They did not find any significant difference in radio flux density between RS CVn systems and Algol-type binaries. Umana *et al.* (1993) also made observations of radio continuum spectra of several Algol-type binaries with the VLA at frequencies of 1.49, 5.0, 8.4, and 14.9 GHz. The radio spectrum of RZ Cas was found to be similar to those of RS CVn binaries. In addition, Umana *et al.* (1999) showed that the RZ Cas system shows variability in the radio flux that cannot be attributed to geometrical effects such as a partial eclipse of the radio source.

A number of optical observations have been made of this object, particularly photometry of the primary eclipse. Several authors argued about the nature of the primary minimum, as to whether it is a partial minimum or a total one (e.g., Chambliss 1976; Nowak and Piotrowski 1982; Olson 1982; Arganbright *et al.* 1988; Hegedus 1989; Nakamura *et al.* 1991a, b; Hegedus *et al.* 1992). This point has been one of the contentious issues on RZ Cas in recent years.

Short-period light variations of RZ Cas were reported by several authors (Olson 1982; Edwin and Gears 1992; Davis and Balonek 1996). Oshima *et al.* (1998, 2001) were the first to show that this variability is caused by oscillations of the primary with a dominant frequency of 64.2 cd$^{-1}$.

In many cases, the period variation was explained assuming RZ Cas to be a triple, quadruple, or quintuple system (e.g. Panchatrasam 1981; Abhyankar and Ballabh 1983; Khaliullina 1987). Hegedus *et al.* (1992) proposed period variations, which they supposed was caused by RZ Cas being even a sextuple system! But according to the empirical stability criteria for hierarchical systems (Roy 1979, 1982) and the stability constraints for planetary systems (Graziani and Black 1981) the systems containing more than three components are unstable.

Conflicting reports on the changing period and other phenomena of RZ Cas warranted a closer study the O–C behavior of this active Algol binary.

## 2. O–C diagram analysis

We decided to plot the O–C curve for RZ Cassiopeia using AAVSO data (Baldwin and Samolyk 1993, 1995, 1997, 1999, 2000, 2002, 2003). The light-elements used are from the *General Catalogue of Variable Stars* (Kholopov *et al.*, 1985). There are 357 times of minimum available for about 30 years. The observations span JD 2442412 (December, 30, 1974) to 2452678 (February, 7, 2003).

The curve in Figure 1 is a parabola fitted to the O–C points by least squares. The standard deviation is 0.00421. A square term for a parabola of $1.002 \times 10^{-9} \pm 4.04928 \times 10^{-11}$ was found. This implies that $\dot{P} = (dP/dt)$ is equal to $2.004 \times 10^{-9} \pm 8.09856 \times 10^{-11}$.

The new parabolic elements are:

$$JD_{(min)} = 2443200.3063 + 1.195247\,E + 1.002 \times 10^{-9}\,E^2 \qquad (1)$$

The long-term trend in Figure 1 indicates that the period has been increasing since JD 2442412. An explanation of the parabolic curve of the RZ Cas O–C could lie in the slow changing of the system's general center of mass which is caused, perhaps, by mass exchange between the components. Considering this trend being the result of a mass transfer, we tried to estimate the rate $\dot{M}$ of the mass transfer. Kwee (1958) determined a relationship for the change in the orbital period assuming total orbital angular momentum is conserved while mass is transferred from one star to another.

Reproducing equation (5) of Kwee (1958),

$$\frac{\Delta P}{P} = 3\left(\frac{m_s}{m_p} - 1\right)\frac{dm_s}{m_s} \qquad (2),$$

where $\Delta P$ is the change in the period, P, $m_s$ is the mass of the secondary component of the system, $m_p$ is the mass of the primary star, and $dm_s$ is the change in mass of $m_s$, we tried to establish a value of the $\dot{M}$ parameter for RZ Cas.

Using the parameters for the RZ Cas system determined by Giuricin *et al.* (1983), the mass of the secondary $m_s$ can be set to 0.65 $M_{\odot}$ and the mass $m_p$ to 1.9 $M_{\odot}$. In this way, equation (2) gives us a mass transfer rate estimate of:

$$M = 2.01 \times 10^{-7} M_{\odot} \times yr^{-1} \qquad (3).$$

Also we can see that upon parabola fitting some oscillations are superimposed. We extract the parabolic trend from O–C to see residuals. Such residuals show several alternate oscillations (waves) as well as some random, more frequent, sharp, and numerous fluctuations. Alternate oscillations shown in Figure 2 are marked by a line. It is not meant to be an approximation. It is just that the line might indicate the presence of such waves.

We apply the Stellingwerf method for retrieving any periodicity in residuals distribution. In case of presence of additional body, these waves would show strict periodicity (Rovithis-Livaniou 2001). To generate the Stellingwerf periodogram (Stellingwerf 1978) we use ISDA (Pelt 1992). As shown in Figure 3, the significant peaks correspond to the values of the lengths of alternate waves (for example, 20.5 years). So, these peaks could not be interpreted as a value of the periods because the analyzed time scale doesn't much exceed the found lengths of alternate waves. Analysis by the Stellingwerf method for searching the periodicity on smaller time scales didn't reveal any other significant peaks on the periodogram. That is why we believe these alternate oscillations are not periodic.

One may suspect that sinusoidal period changes are also possible. The description of an O−C diagram of an eclipsing binary using a sinusoidal term can be physically supported. In such a case, the existence of a third body could be responsible for a sinusoidal variation from the well-known light-time effect (Borkovits and Hegedus 1996). Future observations may help to clarify this fundamental point and related problems. But apsidal motion, which involves a change in the orientation of the binary's major axis, is an unlikely mechanism because close binaries posses circular orbits, while this phenomenon only occurs in systems having large eccentricities. If apsidal motion was present, the times for secondary and primary minima would be shifted in opposite directions, but this effect is rarely seen and does not occur in Algols.

Finally, Hall (1989) wrote that, for Algols that manifest alternating period increases and decreases, the spectral types of the secondaries are always in the range of late F to K class. This is in good agreement with determination of the RZ Cas secondary spectral class as K0IV (Narusawa *et al.* 1994).

Another possible explanation could be the development of magnetic activity cycles in one or both components (Applegate 1992). Zavala *et al.* (2002) give the WW Cyg O−C, which is similar to that of RZ Cas. They propose their explanation of such O−C using magnetic activity theory. So, we believe that the reason of such O−C changes would be common.

The large scatter in O–C residuals (near $\pm 0.009$d = 13 min) is also very interesting. Perhaps it is caused by a short-periodic pulsation of the RZ Cas primary component, which is deforming the shape of the primary eclipse. Figure 4 shows schematic lightcurves with extrema indicated in 4 different cases of primary eclipse, distorted by pulsation. This suggestion is compatible with the 22.4-minute period Oshima *et al.* (1998) give for the delta Scuti oscillations. Reports on flat-bottomed eclipses lasting up to 22 minutes (e.g. Arganbright *et al.* 1988) are also agreeable with this hypothesis. Following the work of Lehmann and Mkrtichian (2004), the rapid photometric oscillations of the primary component were monoperiodic from 1997 to October 2000 with a frequency of 64.2 $cd^{-1}$ and a semi-amplitude of about 9 mmag; but in 2001 the spectrum of the pulsations became multiperiodic and the photometric amplitude decreased by one order of magnitude.

## 3. Conclusion

An O–C diagram covering about 30 years was plotted for RZ Cas. Polynomial fitting of the second degree was used. Was found square term for RZ Cas light-elements: $1.002 \times 10\text{-}9 \pm 4.04928 \times 10\text{–}11$. $\dot{P}$ is equal to $2.004 \times 10^{-9} \pm 8.09856 \times 10^{-11}$. The new parabolic elements are: $JD_{(min)} = 2443200.3063 + 1.195247\,E + 1.002 \times 10^{-9}\,E^2$. Analysis of O–C gave us an opportunity to establish the value of $\dot{M}$ to be equal to $2.01 \times 10^{-7}\ M_{\odot} \times yr^{-1}$. Application of the Stellingwerf method for analysis of the residuals after extracting parabola did not reveal any strict and certain periodicy in the residuals.

Algol-type stars change their orbital periods through mass-exchange processes that change the center of mass of the system. RZ Cas, perhaps, is one of them. Greater attention should to be paid to the fact that the short-periodic pulsations of the primary component have a great influence on the shape of the eclipse and can cause pseudochanging (spurious changing) of orbital period.

## 4. Acknowledgements

We are indebted to the many AAVSO observers, amateur and professional, who amassed the wealth of data on times of RZ Cas minima. This research has made use of the SIMBAD database, operated at CDS, Strasbourg, France, and the NASA Astrophysics Data System Abstract Service.

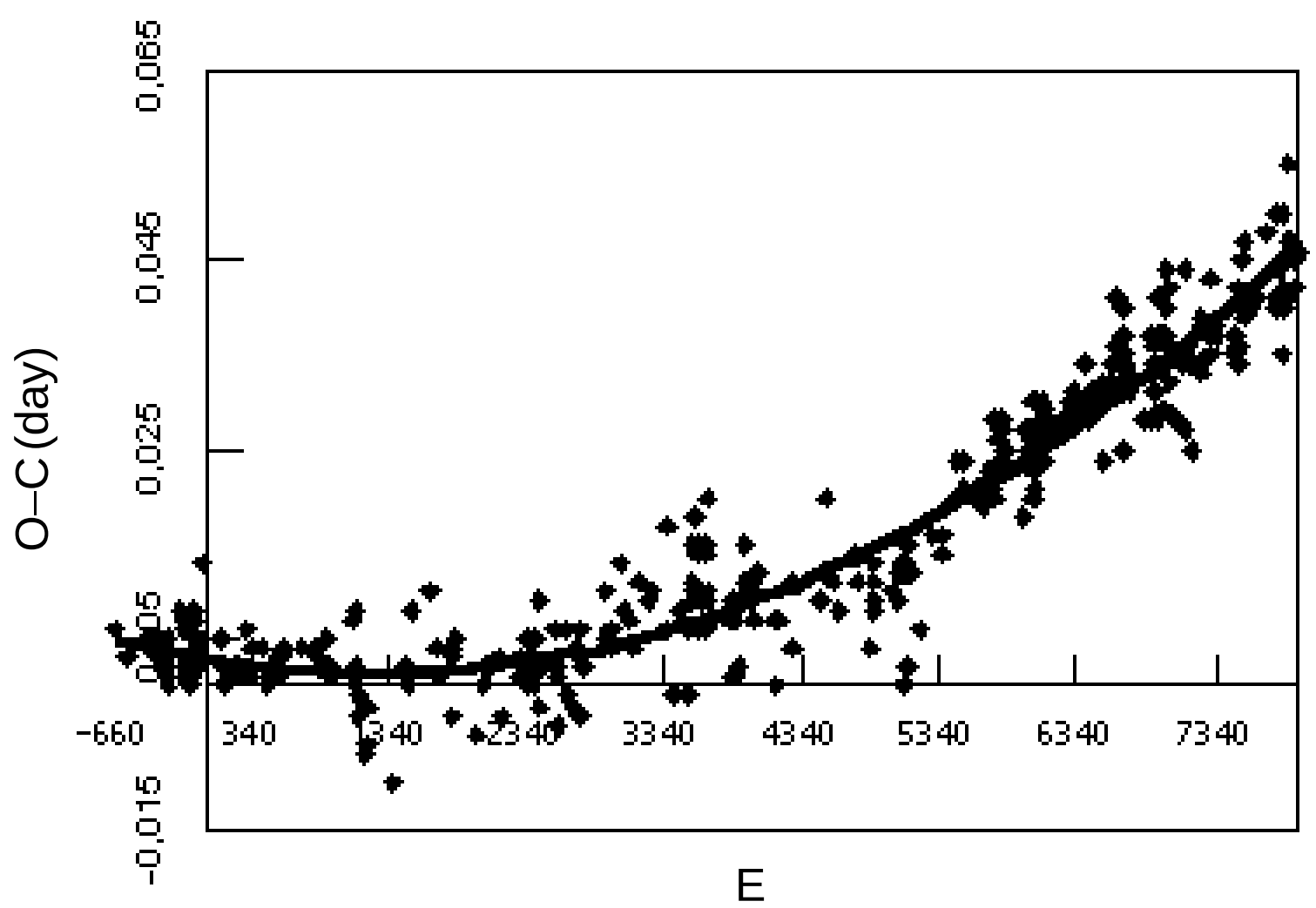

Figure 1. O–C diagram for RZ Cas.

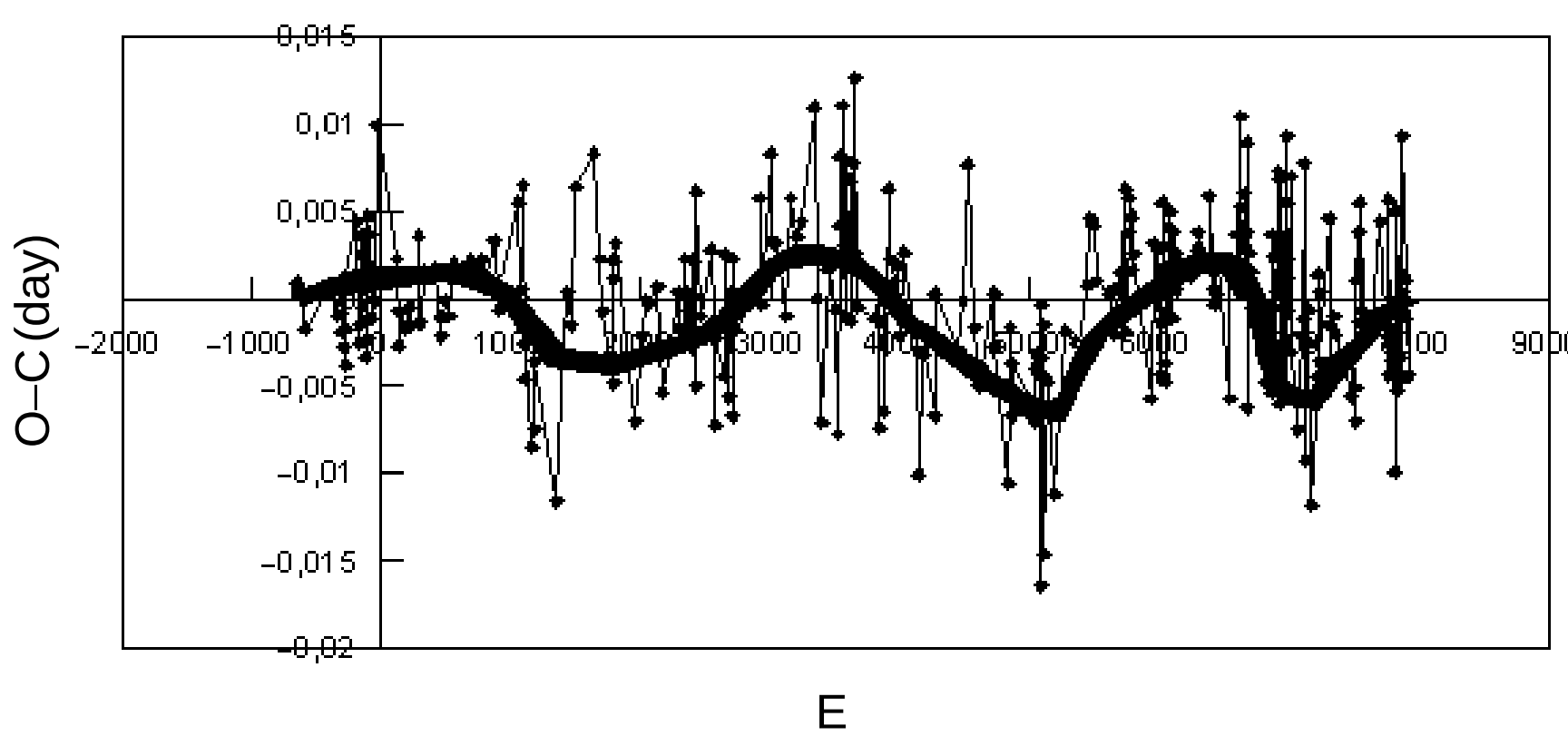

Figure 2. Residuals after excluding parabola fitting. Alternate oscillations are indicated by the hand-drawn line.

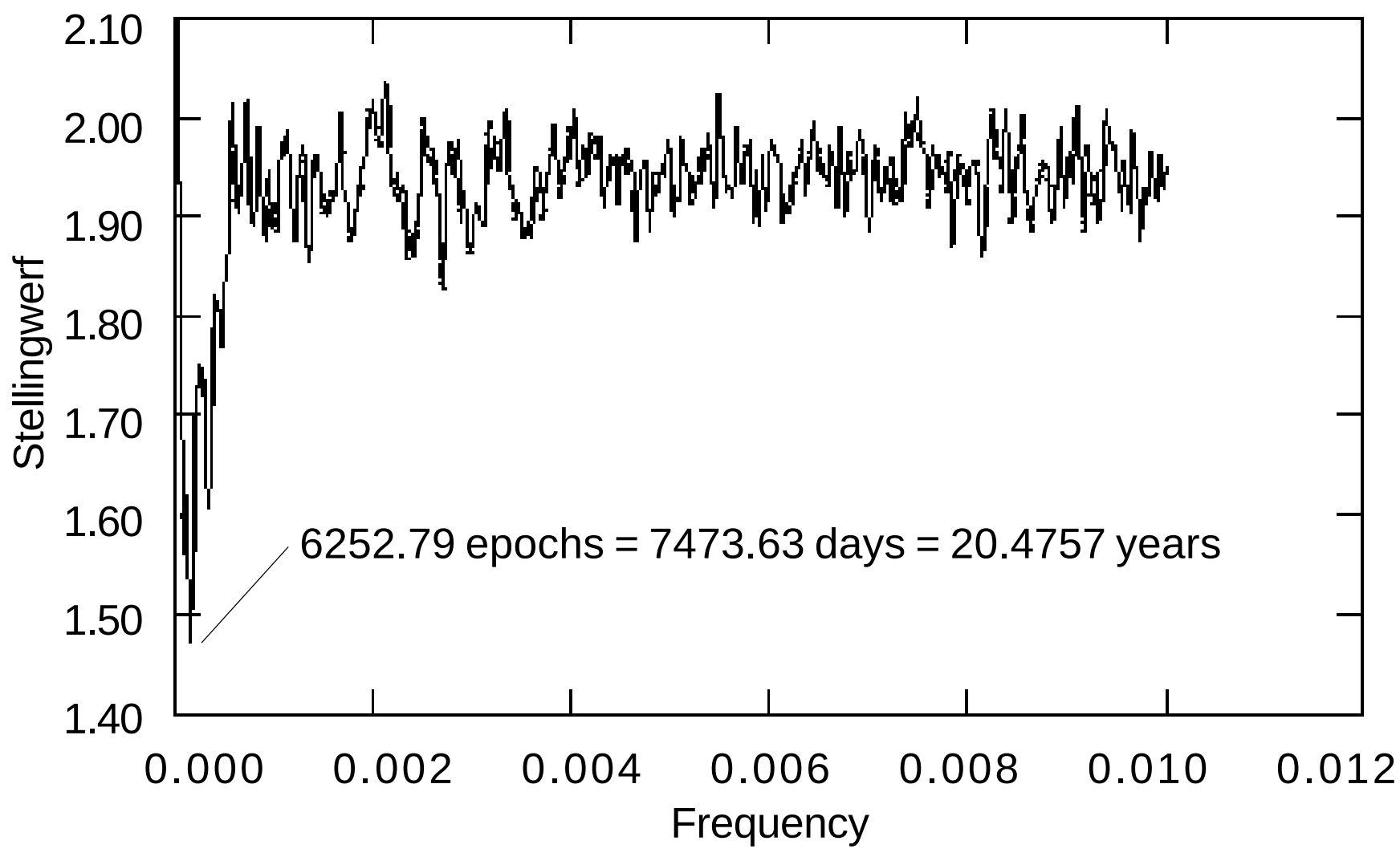

Figure 3. Stellingwerf periodogram for residuals after excluding parabola fitting.

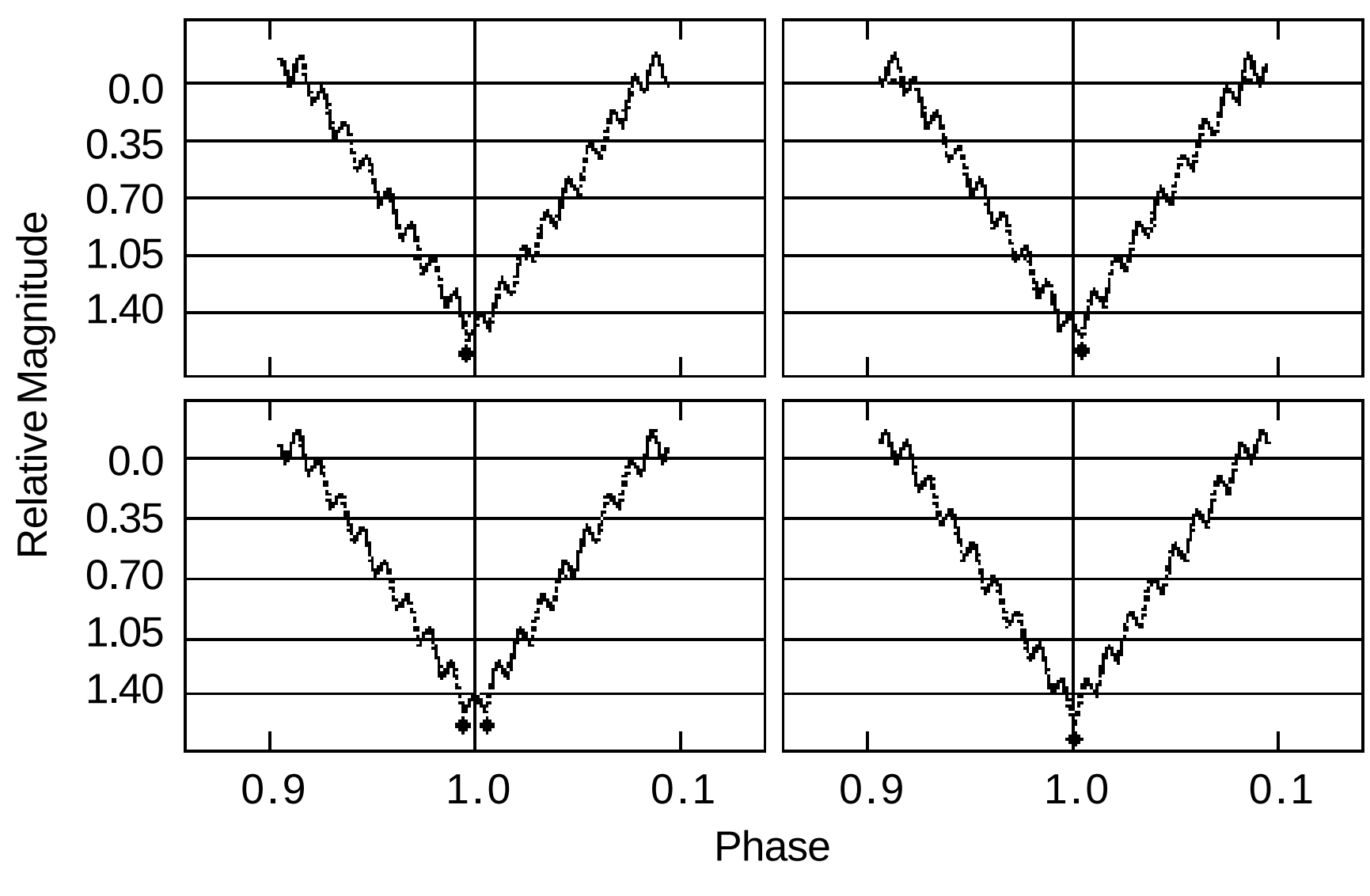

Figure 4. Schematic lightcurve of eclipse, distorted by pulsations of the primary component.